\documentclass[aps,prd,superscriptaddress,nofootinbib,amsmath,amsfonts,preprintnumbers,groupedaddress,showpacs,10pt,english]{revtex4-2}
\usepackage{amsmath} 
\usepackage{booktabs}
\usepackage{placeins}
\usepackage{amssymb}
\usepackage{babel}
\usepackage{wrapfig}
\usepackage{cancel}
\usepackage{float}

\newcommand{\Mp}{M_{\mathrm{P}}}
\newcommand{\ar}{\bar{\alpha}}
\usepackage{relsize,exscale}
\usepackage{array}
\makeatletter

\usepackage{array,multirow,graphicx}
\usepackage{soul}
\usepackage{dcolumn}
\usepackage{newlfont}
\usepackage{bm}
\usepackage[colorlinks,citecolor=blue,urlcolor=blue,linkcolor=blue]{hyperref}
\usepackage[figtopcap]{subfigure}
\usepackage{color}
\usepackage{verbatim}

\def\be{\begin{align}}
\def\ee{\end{align}}
\def\bea{\begin{eqnarray}}
\def\eea{\end{eqnarray}}
\def\bal{\begin{align}}
\def\eal{\end{align}}

\usepackage{scalerel}
\usepackage{tikz}
\usetikzlibrary{svg.path}
\definecolor{orcidlogocol}{HTML}{A6CE39}
\tikzset{
 orcidlogo/.pic={
 \fill[orcidlogocol] svg{M256,128c0,70.7-57.3,128-128,128C57.3,256,0,198.7,0,128C0,57.3,57.3,0,128,0C198.7,0,256,57.3,256,128z};
 \fill[white] svg{M86.3,186.2H70.9V79.1h15.4v48.4V186.2z}
 svg{M108.9,79.1h41.6c39.6,0,57,28.3,57,53.6c0,27.5-21.5,53.6-56.8,53.6h-41.8V79.1z M124.3,172.4h24.5c34.9,0,42.9-26.5,42.9-39.7c0-21.5-13.7-39.7-43.7-39.7h-23.7V172.4z}
 svg{M88.7,56.8c0,5.5-4.5,10.1-10.1,10.1c-5.6,0-10.1-4.6-10.1-10.1c0-5.6,4.5-10.1,10.1-10.1C84.2,46.7,88.7,51.3,88.7,56.8z};}}
\newcommand\orcid[1]{\href{https://orcid.org/#1}{\mbox{\scalerel*{
\begin{tikzpicture}[yscale=-1,transform shape]
\pic{orcidlogo};
\end{tikzpicture}
}{|}}}}
\graphicspath{{./}{Figs/}}
\def\Vec#1{\mbox{\boldmath $#1$}}
\begin{document}


\title{Scalaron initial conditions and regular-branch response\\in quadratic scalar--vector--tensor cosmology}
\author{G.~G.~N.~Nashed$^{1}$}\email{nashed@bue.edu.eg}
\author{Kazuharu Bamba$^{2}$}\email{bamba@sss.fukushima-u.ac.jp}
\affiliation{Centre for Theoretical Physics, The British University in Egypt, P.O. Box 43, El Sherouk City, Cairo 11837, Egypt\\
$^{2}$Faculty of Symbiotic Systems Science, Fukushima University, Fukushima 960-1296, Japan}

\begin{abstract}
We formulate the metric quadratic-curvature representation of a scalar--vector--tensor dark-energy background, keeping the nonzero temporal-vector solution used in the general-relativistic theory. We show that the additional curvature degree of freedom can influence on the choice of initial condition. We find that changing the initial curvature or its derivative alters the effective pattern of the phantom-divide crossing. We also analyze the rapid homogeneous curvature oscillation and obtain its leading frequency and damping in the presence of the temporal vector. As a result, we demonstrate that for the branch obtained by omitting the free scalaron mode, order reduction gives only a small normalized change in distance over the reference coupling interval used as the test of the consistency. Furthermore, we perform that the information diagnostic cannot be interpreted as an observational bound on the coupling term of the quadratic curvature. 
\end{abstract}
\maketitle

\section{Introduction}\label{sec:introduction}

The scalar--vector--tensor (SVT) cosmological model investigated by
Roy \textit{et al.}~\cite{Roy} provides a framework in which a canonical scalar
field and a temporal vector field can produce an effective dark-energy
equation of state that crosses the phantom divide. Their analysis was
carried out within general relativity (GR). In the present work, we
retain the scalar--vector matter sector of that model while extending
the gravitational action to metric $f(R)$ gravity~\cite{DeFelice}.
This modification introduces an additional scalar degree of freedom
associated with the Ricci curvature. Consequently, the cosmological
evolution depends not only on the conventional background variables
but also on the initial values of $R$ and $\dot R$, which must be
specified consistently with the gravitational constraints. Depending
on these initial conditions, the resulting solutions may remain close
to their GR counterparts or exhibit appreciable curvature
oscillations. Understanding this dependence is one of the principal
objectives of the present investigation.

The theoretical foundations and cosmological implications of metric
$f(R)$ gravity have been extensively discussed in the literature.
Sotiriou and Faraoni~\cite{Sotiriou2008} presented a comprehensive
account of its formulation and physical properties, while
Capozziello and De Laurentis~\cite{Capozziello2011} examined its
gravitational and astrophysical applications. Broader perspectives
on modified theories of gravity and their cosmological consequences
were provided by Clifton \textit{et al.}~\cite{Clifton2011} and Nojiri and
Odintsov~\cite{Nojiri2010}. Further developments, including
applications to inflation, bouncing cosmologies, and late-time
acceleration, were reviewed by Nojiri, Odintsov, and
Oikonomou~\cite{Nojiri2017}. These studies demonstrate the variety
of cosmological phenomena that can arise from modifications of the
gravitational action. They also motivate an examination of how an
additional curvature degree of freedom affects a cosmological
background containing interacting scalar and vector fields.

The construction of viable metric $f(R)$ cosmologies requires
particular attention to their background evolution and stability
properties. Representative models compatible with local
gravitational constraints were developed by Hu and Sawicki and
Starobinsky~\cite{HuSawicki2007,Starobinsky2007}. The cosmological
phase-space structure and conditions for acceptable evolutionary
histories were investigated in
Refs.~\cite{Amendola2007,AmendolaPolarski2007}, whereas the behaviour
and stability of high-curvature solutions were addressed in
Refs.~\cite{SawickiHu2007,DolgovKawasaki2003,Faraoni2006}.
Earlier modifications of the gravitational action and their
connections with scalar--tensor representations were explored in
Refs.~\cite{Carroll2004,Chiba2003,Sotiriou2006,NojiriOdintsov2006}.
An acceptable homogeneous background, however, does not necessarily
guarantee perturbative stability or consistency with the observed
growth of cosmological structures, as emphasized in
Refs.~\cite{SongHuSawicki2007,Bean2007,PogosianSilvestri2008}.
Alternative reconstruction procedures and representative functional
forms have also been examined in
Refs.~\cite{Dunsby2010,Linder2009}. Although these investigations
provide important guidance, their conclusions cannot be transferred
directly to the quadratic SVT model considered here, since the
underlying gravitational actions and matter sectors are generally
different.

In this study, we adopt the quadratic gravitational Lagrangian
$f(R)=R+\alpha R^2$. The importance of the quadratic curvature term
was established in the inflationary scenario proposed by
Starobinsky~\cite{Starobinsky1980}. Our analysis instead concerns
its influence on late-time cosmological evolution in the presence
of a scalar--vector background. Accordingly, the parameter $\alpha$
is treated independently of the value associated with the
inflationary application. For $\alpha>0$, the trace equation
supports an oscillatory scalaron mode whose amplitude and phase
depend on the initial curvature data. This dependence is
particularly relevant when comparing solutions of the full
higher-order field equations with perturbative solutions connected
continuously to GR.

A useful distinction arises between the complete curvature dynamics
and the order-reduced approximation. The latter selects a
perturbative solution in which the freely oscillating scalaron
contribution is excluded. Nevertheless, matching the initial
curvature data to a finite perturbative order does not necessarily
suppress this mode completely. The reduction-of-order approach to
higher-derivative gravitational equations was developed by
Simon~\cite{Simon1990} and Parker and Simon~\cite{ParkerSimon1993}.
A related procedure was subsequently applied to
$f(R,\mathcal G)$ cosmology by Barros \textit{et al.}~\cite{Barros}.
The physical significance of scalaron oscillations in other
cosmological circumstances was discussed in Ref.~\cite{Shtanov}.
It is therefore essential to distinguish the properties of the
GR-connected perturbative branch from those of solutions in which
the independent curvature mode is excited. Reduction of order
provides a controlled perturbative description but does not remove
the additional scalar degree of freedom from the underlying metric
$f(R)$ theory.

The dependence of cosmological evolution on the scalaron mode has
also been examined in several related studies. Tsujikawa 
\textit{et al.}~\cite{Tsujikawa2008} discussed the oscillatory behaviour
of curvature degrees of freedom in viable $f(R)$ models.
Motohashi \textit{et al.}~\cite{Motohashi2011} investigated repeated
crossings of the effective phantom divide during the approach
towards a de Sitter configuration. Related oscillatory
dark-energy behaviour was considered in other modified-gravity
models~\cite{Nojiri2025}. The nonlinear evolution of curvature
may introduce additional complications. Appleby, Battye, and
Starobinsky~\cite{Appleby2010} demonstrated that a quadratic
curvature correction can regularize certain high-curvature
pathologies, while Frolov~\cite{Frolov2008} identified an
accessible curvature singularity in a class of infrared-modified
gravity models. These results underline the importance of
examining the allowed solution branches and their domains of
validity. They do not, however, establish the presence or absence
of corresponding singularities in the SVT extension investigated
in the present work.

The vector sector of the model is related to developments in
generalized Proca theories, which incorporate derivative
self-interactions of massive vector fields. The theoretical
construction of these interactions and their cosmological
applications were discussed in
Refs.~\cite{Heisenberg2014,Tasinato2014,BeltranHeisenberg2016,Proca2016}.
Their consequences for effective gravitational couplings were
investigated in Ref.~\cite{DeFeliceProca2016}, while a broader
review of vector--tensor modified gravity was presented in
Ref.~\cite{HeisenbergReview2019}. The scalar sector is connected
with the general scalar--tensor framework introduced by
Horndeski~\cite{Horndeski1974} and its subsequent cosmological
developments~\cite{Kobayashi2011,KaseTsujikawaReview2019}.
These formulations provide the theoretical background for
constructing interacting scalar--vector cosmologies. Nevertheless,
their perturbative stability conditions cannot be adopted without
modification when the gravitational action contains an additional
metric-$f(R)$ scalar degree of freedom.

The temporal-vector tracker employed in our analysis is motivated
by the generalized Proca cosmological construction~\cite{Proca2016}.
A more general framework describing interactions between scalar
and vector fields was developed by Heisenberg~\cite{Heisenberg2018}.
The corresponding cosmological dynamics were investigated by
Heisenberg, Kase, and Tsujikawa~\cite{SVTCosmology2018}.
Kase and Tsujikawa~\cite{SVTDarkEnergy2018} further examined
SVT dark-energy models at both the background and perturbative
levels, including the influence of matter on scalar propagation
and the effects of the vector sector on structure growth and
gravitational lensing. A systematic gauge-ready treatment of the
coupled perturbations was developed in Ref.~\cite{GaugeReady2018}.
These results are particularly relevant because the introduction
of the quadratic curvature term changes the number of propagating
scalar degrees of freedom. Consequently, the stability of the
extended theory must be established from its complete perturbation
system rather than inferred from the properties of the original
SVT model.

Against this background, the present investigation addresses two
closely related aspects of quadratic $f(R)$--SVT cosmology. The
first concerns the sensitivity of late-time cosmological evolution
to the initial curvature and its time derivative. We examine
different curvature initial states, each satisfying the Friedmann
constraint, and determine how their subsequent evolution compares
with the GR background. The numerical solutions are independently
checked using two integration methods. The second aspect concerns
the influence of the temporal vector field on the rapidly
oscillating homogeneous scalaron mode. In particular, we examine
its frequency, damping, and amplitude on the tracker branch
characterized by $\rho_\chi\propto H^{-2s}$. The local
short-timescale description is compared with direct numerical
evolution to assess its range of applicability.

We also investigate the normalized cosmological distance response
obtained from the GR-connected order-reduced solution. Supernova
and baryon acoustic oscillation (BAO) covariance matrices are
used to quantify the sensitivity of this response to the
quadratic curvature correction. This analysis is intended as a
diagnostic of the perturbative approximation rather than a
likelihood determination of the parameters of the full
$f(R)$--SVT theory. The reference cosmological parameters are
taken from the marginal mean values reported by Roy
et al.~\cite{Roy}; they are not obtained from a new joint
best-fit analysis. Likewise, the homogeneous scalaron calculation
does not establish the absence of ghosts or gradient instabilities
in the complete coupled perturbation system. The purpose is to
identify the dependence of the background and curvature-mode
dynamics on the chosen initial conditions and on the temporal
vector contribution, while keeping the physical interpretation
within the scope of the calculations performed.

The remainder of the paper is organized as follows.
In Sec.~\ref{sec:action}, we present the gravitational action,
derive the background field equations, and establish the
corresponding constraint identity. Section~\ref{sec:benchmarks}
examines the numerical evolution of representative late-time
cosmological solutions, emphasizing their dependence on the
initial curvature conditions. In Sec.~\ref{sec:response}, we
derive the order-reduced response connected to the GR background.
Section~\ref{sec:fast} is devoted to the rapidly oscillating
homogeneous curvature mode and its behaviour in the
scalar--vector background. The cosmological distance response
and associated information diagnostics are presented in
Sec.~\ref{sec:distance}. We discuss the interpretation and
limitations of the results in Sec.~\ref{sec:discussion}, followed
by our conclusions in Sec.~\ref{sec:conclusions}. Further details
of the numerical validation, initial-data matching, and
computational procedures are given in 
Appendices~\ref{app:validation} and \ref{app:initialdata}.
Throughout this work, the expression ``exact evolution'' refers
to numerical solutions of the original field equations over the
specified integration intervals. It does not denote an analytical
solution or imply that the resulting evolution has been
established throughout the entire cosmological history. 
Throughout this paper, we use the metric signature of $(-,+,+,+)$. 

\begin{table}[th]
\centering
\caption{Scope of the calculation families.}
\label{tab:scope}
\small
\begin{tabular}{@{}ll@{}}
\hline\hline
\textbf{Construction} & \textbf{Scope and limitations} \\
\hline
Original quadratic equations &
Local checks; no global history. \\
Formal regular response &
First-order checks; rapid mode omitted. \\
Finite-parameter slow closure &
Reduced solution; no global error bound. \\
Transported modes &
Conditional diagnostics; no global validation. \\
\hline
\end{tabular}
\end{table}

\section{Theory and constrained background}\label{sec:action}
We consider the following action, in which only the Einstein--Hilbert term in Eq.~(2.1) of Ref.~\cite{Roy} is replaced, given by

\begin{align}
S={}&\int d^4x\sqrt{-g}
\left[
\frac{M_{\mathrm{P}}^2}{2}f(R)
+\mathcal{F}
+G_2(X)
\right]
\nonumber\\
&+\int d^4x\sqrt{-g}
\left[
G_3(X)\nabla_\mu A^\mu
-\frac{1}{2}(\nabla\phi)^2
-V(\phi)
\right]+S_\mathrm{m} .
\end{align}
Here, $X=-A_\mu A^\mu/2$, $\mathcal F=-F_{\mu\nu}F^{\mu\nu}/4$, $G_2=b_2X^{p_2}$, $G_3=b_3X^{p_3}$, and $V=V_0e^{-\lambda\phi/M_{\mathrm{P}}}$. The curvature sector is varied with respect to the metric (rather than through the Palatini prescription). We write $F=f_{,R}$, reserving $\mathcal F$ for the vector invariant. On the spatially flat FLRW background we take $ds^2=-dt^2+a^2(t)d\boldsymbol x^2$ and the contravariant ansatz $A^\mu=(\chi(t),0,0,0)$, so $A_0=-\chi$, $X=\chi^2/2$, and $F_{\mu\nu}=\nabla_\mu A_\nu-\nabla_\nu A_\mu$ vanishes on the background. We take the nonzero temporal branch with $p\ne0$ and a real choice of the power-law functions. The scalar equation and temporal-vector constraint are unchanged:
\begin{equation}
\ddot\phi+3H\dot\phi+V_{,\phi}=0,\qquad G_{2,X}+3H\chi G_{3,X}=0.
\end{equation}
On this branch $\chi^pH=\mathrm{const}$, with $p=1-2p_2+2p_3$ and $s=p_2/p$, so $\rho_\chi\propto H^{-2s}$. More explicitly, on this branch,
\begin{align}
H\chi^p&=-\frac{b_2p_2\,2^{p_3-p_2}}{3b_3p_3},\\
\rho_\chi&=-G_2=-b_2\left(\frac{\chi^2}{2}\right)^{p_2}.
\end{align}
We require $b_3p_3\ne0$ and choose a real field branch with positive vector density. At background level the calculations fix $s$ and the density amplitude, rather than a unique set of microscopic coefficients and exponents. Separate conservation gives
\begin{equation}
P_\chi=-\rho_\chi+\frac{2s}{3}\rho_\chi\frac{\dot H}{H^2}.
\end{equation}
Matter and radiation continue to satisfy their usual conservation equations. The extra scalaron changes the perturbation problem, however, and the GR conditions alone no longer establish stability of the extended system.

\subsection{General metric \texorpdfstring{$f(R)$}{f(R)} equations}\label{sec:general}
Metric variation gives the following standard field equation~\cite{DeFelice}:
\begin{equation}
FR_{\mu\nu}-\tfrac12fg_{\mu\nu}+(g_{\mu\nu}\Box-\nabla_\mu\nabla_\nu)F=\Mp^{-2}T_{\mu\nu}.
\end{equation}
The stress tensor $T_{\mu\nu}$ contains the matter and radiation contributions together with the two SVT sectors.
We take the flat Friedmann--Lema\^{\i}tre--Robertson--Walker (FLRW) space-time with the metric, given by
\begin{align}
{ds}^2 = -{dt}^2 + a^2(t)d{\Vec{x}}^2\,,
\end{align}
where $a(t)$ is the scale factor.
In the FLRW background, the equations become
\begin{align}
3FH^2&=\frac{\rho}{\Mp^2}+\frac{FR-f}{2}-3H\dot F,\\
-2F\dot H&=\frac{\rho+P}{\Mp^2}+\ddot F-H\dot F,\\
\ddot F+3H\dot F&=\frac{FR-2f+(\rho-3P)/\Mp^2}{3},\\
R&=6(2H^2+\dot H).
\end{align}
If $f_{,RR}\ne0$, the trace can be written as
\begin{align}
\ddot R={}&-3H\dot R-\frac{f_{,RRR}}{f_{,RR}}\dot R^2\nonumber\\
&+\frac{FR-2f+(\rho-3P)/\Mp^2}{3f_{,RR}}.
\end{align}
To specify a solution, we need scalaron initial data as well as the scalar and vector data, all subject to the Friedmann constraint. Imposing a boundary condition at late times does not by itself select a branch that is regular in the radiation era.

\subsection{Explicit quadratic system}\label{sec:quadratic}
For $f=R+\alpha R^2$, we find
\begin{equation}
F=1+2\alpha R,\qquad f_{,RR}=2\alpha,\qquad FR-2f=-R.
\end{equation}
Consequently,
\begin{equation}
\ddot R+3H\dot R+\frac{R-(\rho-3P)/\Mp^2}{6\alpha}=0.
\end{equation}
The curvature-sector mass coefficient is $m_R^2=1/(6\alpha)$. Taking $\alpha$ positive and requiring $F>0$ satisfies these basic curvature-sector conditions. Stability of the coupled scalar and vector perturbations requires further checks.
Let $N=\ln a$, $E=H/H_{\mathrm{ref}}$, $r=R/H_{\mathrm{ref}}^2$, $q=\dot R/H_{\mathrm{ref}}^3$, $\ar=\alpha H_{\mathrm{ref}}^2$, $v=\dot\phi/(\Mp H_{\mathrm{ref}})$, $\varphi=\phi/\Mp$, and $U=V/(3\Mp^2H_{\mathrm{ref}}^2)$. Define $d_i=\rho_i/(3\Mp^2H_{\mathrm{ref}}^2)$.
We then obtain
\begin{align}
\epsilon&=\frac{r}{6E^2}-2,\quad d_\mathrm{m} =\Omega_{\mathrm{m}, \mathrm{ref}}e^{-3N},\nonumber\\
d_\mathrm{r}&=\Omega_{\mathrm{r},\mathrm{ref}}e^{-4N},\quad d_\chi=\Omega_{\chi,\mathrm{ref}}E^{-2s},\\
\mathcal T&=3d_\mathrm{m} +12d_\chi-6sd_\chi\epsilon-v^2+12U.
\end{align}
The six-dimensional system is
\begin{align}
E_N&=E\epsilon,\qquad r_N=q/E,\\
q_N&=-3q-\frac{r-\mathcal T}{6\ar E},\\
\varphi_N&=v/E,\quad v_N=-3v+3\lambda U/E,\\
U_N&=-\lambda Uv/E.
\end{align}
Its dimensionless constraint is
\begin{align}
\mathcal C={}&E^2(1+2\ar r)- d_\mathrm{m} -d_\mathrm{r}-d_\chi\nonumber\\
&-\frac{v^2}{6}-U-\frac{\ar r^2}{6}+2\ar Eq=0.
\end{align}
This ODE system becomes singular as $\alpha\to0$, so we obtain GR from its own equations. A smooth GR limit requires a prescription for the homogeneous scalaron mode as well as a small $\alpha$.

\subsection{Constraint propagation and effective crossings}\label{sec:constraint}
Substituting the evolution equations and the fluid conservation laws into the derivative of the constraint gives
\begin{equation}
\mathcal C_N=-4\mathcal C.
\end{equation}
Thus an initially vanishing constraint continues to vanish under the evolution equations. For comparison, we write the dimensionless residual of the acceleration equation as
\begin{align}
\mathcal A={}&-2(1+2\ar r)E^2\epsilon\nonumber\\
&-\left[3d_\mathrm{m} +4d_\mathrm{r} +v^2+2sd_\chi\epsilon+2\ar E(q_N-q)\right].
\end{align}
Along solutions of the evolution equations, $\mathcal A=4\mathcal C$. The two residuals are algebraically dependent and supply only one constraint check. We use the agreement between distinct integration algorithms as a separate numerical test.

For a geometric split with constant $\Mp$,
\begin{align}
\frac{\rho_{\mathrm{eff}}}{\Mp^2H_{\mathrm{ref}}^2}&=3(E^2-d_\mathrm{m}-d_\mathrm{r}),\label{eq:effrho}\\
\frac{\rho_{\mathrm{eff}}+P_{\mathrm{eff}}}{\Mp^2H_{\mathrm{ref}}^2}&=-2E^2\epsilon-3d_\mathrm{m}-4d_\mathrm{r},\\
w_{\mathrm{eff}}&=-1+\frac{\rho_{\mathrm{eff}}+P_{\mathrm{eff}}}{\rho_{\mathrm{eff}}}.
\end{align}
For this definition of the effective fluid, an enthalpy sign change represents a crossing only if the density there is finite and nonzero. This numerical agreement applies to the chosen initial state; other scalaron branches can have different histories.

\section{Initial data and exact late-time benchmarks}\label{sec:benchmarks}
We use the marginal means in Table~II of Ref.~\cite{Roy}, not a published joint best fit: $H_{\mathrm{ref}}=67.0736\,\mathrm{km\,s^{-1}\,Mpc^{-1}}$, $\Omega_{m,\mathrm{ref}}=0.311773$, $\Omega_{\mathrm{b},\mathrm{ref}}=0.0497458$, $f_\chi=0.421421$, $s_\chi=0.163257$, and $\lambda=1.797894$, with $s=s_\chi/f_\chi$. The reference vector amplitude is $\Omega_{\chi,\mathrm{ref}}=f_\chi(1-\Omega_{\mathrm{m},\mathrm{ref}}-\Omega_{\mathrm{r},\mathrm{ref}})$.
\begin{equation}
\Omega_{\mathrm{r},\mathrm{ref}}=\frac{2.4728\times10^{-5}}{h_{\mathrm{ref}}^2}
\left(1+0.22710731766\,N_{\mathrm{eff}}\right),
\end{equation}
with $h_{\mathrm{ref}}=H_{\mathrm{ref}}/(100\,\mathrm{km\,s^{-1}\,Mpc^{-1}})$ and $N_{\mathrm{eff}}=3.046$, giving $\Omega_{r,\mathrm{ref}}=9.2988\times10^{-5}$. This fixes a photon-plus-relativistic-neutrino reference fluid; it does not compute quadratic recombination or the acoustic scale.
The benchmarks start at $N_i=-\ln4$ ($z_i=3$). We take $E_i=E_g(N_i)$ and $v_i=v_g(N_i)$, with $\varphi_i=0$. We initialize the curvature velocity with its GR reference value, $q_i=q_g(N_i)$. For a trial $r_i$, impose the Friedmann constraint through
\begin{align}
U_i(r_i)={}&E_i^2(1+2\ar r_i)-d_{\mathrm{m},i}-d_{\mathrm{r},i}-d_{\chi,i}\nonumber\\
&-v_i^2/6-\ar r_i^2/6+2\ar E_iq_i.
\label{eq:initialU}
\end{align}
We then choose the root connected to $r_g(N_i)$ of
\begin{equation}
r_i-\mathcal T_i[r_i,U_i(r_i)]
+6\ar E_i[(q_g)_{N,i}+3q_i]=0.
\label{eq:curvaturematch}
\end{equation}
For the benchmark data, we evaluate $q_g=E_g\partial_Nr_g$ with a centered step of $2\times10^{-5}$. We use a centered step of $10^{-4}$ for $(q_g)_N$ and search for the root within $r_g(N_i)\pm30$. The prescription matches the curvature acceleration approximately, to finite order, and can leave a free scalaron component. We solve a separate initial-value problem for each coupling listed in Table~\ref{tab:initialvalues}; the release supplies the starting values at full precision. The present value of $H$ follows from each integration.
In particular, we have
\begin{equation}
H_0=H_{\mathrm{ref}}E(0),\quad
\Omega_{\mathrm{m},0}=\frac{\Omega_{\mathrm{m},\mathrm{ref}}}{E(0)^2},\quad
\alpha H_0^2=\ar E(0)^2.
\end{equation}
Reference normalization and present-day normalization coincide only when the endpoint has the reference value. For the GR solution used here, $E_g(0)=1$, $w_{\mathrm{eff},g}(0)=-0.884109$, and the crossing redshift on this interval is approximately $1.1433$.

\begin{table}[t]
\centering
\caption{Late-time comparison on $0\le z\le3$ using independent integration methods. For each initial state, $n_c$ records the effective crossing count.}\label{tab:benchmarks}
\begin{tabular}{@{}cccc@{}}
\hline\hline
$\ar$&$E(0)$&$\max|\Delta H|/H$&$n_c$\\
\hline
$10^{-5}$&0.990045&$1.31\times10^{-10}$&2\\
$3\times10^{-5}$&0.968213&$2.27\times10^{-10}$&8\\
$10^{-4}$&0.857691&$4.7\times10^{-10}$&5\\
\hline
\end{tabular}
\end{table}

\begin{figure}[th]
\centering
\includegraphics[width=\textwidth]{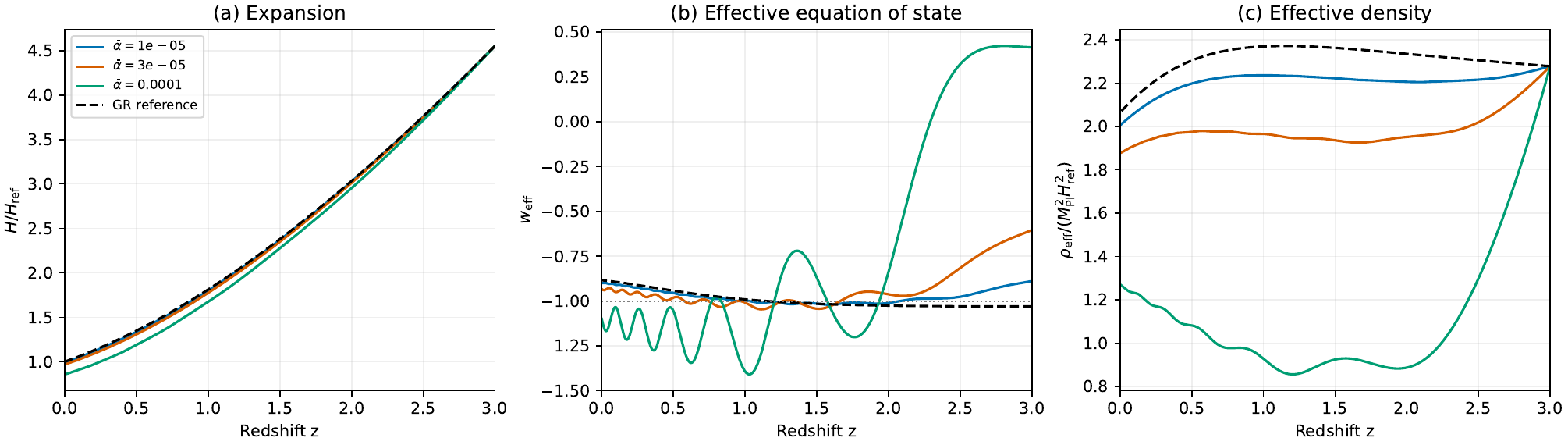}
\caption{Evolution of the quadratic benchmarks and the GR reference. Every example starts at the same $H$ at $z=3$, while the integrations give different $H_0$ values. Both numerical methods resolve the oscillations shown in panel (b). Panels (a)--(c) compare background trajectories; no posterior probabilities are attached to them.}\label{fig:benchmarks}
\end{figure}

The benchmark histories in Table~\ref{tab:benchmarks} and Fig.~\ref{fig:benchmarks} were computed with Radau and checked with BDF at tighter tolerances. Their values of $w_{\mathrm{eff}}$ differ by less than $3.0\times10^{-8}$. In the three baseline cases, the sampled effective density stays positive, reaching minima of about 2.005, 1.877, and 0.856, respectively, in units of $\Mp^2H_{\mathrm{ref}}^2$. The solvers locate each crossing within $1.5\times10^{-7}$ in redshift. We therefore find consistent crossing locations for these specified initial states.

\begin{table}[th]
\centering\small
\caption{Starting data for the original-equation benchmarks at $N_i=-\ln4$, with $\varphi_i=0$ throughout. We give rounded values here and retain full precision in the numerical release.}\label{tab:initialvalues}
\begin{tabular}{@{}cccccc@{}}\hline\hline
$\ar$&$E_i$&$r_i$&$q_i$&$v_i$&$U_i$\\\hline
$10^{-5}$&4.553739834&68.212283935&-813.876980243&0.177858730&0.610812145\\
$3\times10^{-5}$&4.553739834&66.268597216&-813.876980243&0.177858730&0.502523175\\
$10^{-4}$&4.553739834&59.309969563&-813.876980243&0.177858730&0.110513049\\
\hline\end{tabular}
\end{table}

\subsection{Dependence on the initial curvature data}\label{sec:initial}
For the local sensitivity test, we add $\delta r_i=\pm0.01\max(|r_i|,1)$ to $r_i$ in one set of runs and $\delta q_i=\pm0.01\max(|q_i|,1)$ to $q_i$ in another. The starting $E$ and $v$ are held fixed, while $U_i$ is recalculated from the exact Friedmann constraint. For $\ar=10^{-5}$, increasing and decreasing $r_i$ gives 16 and 14 crossings, respectively, against 2 for the baseline. The two changes in $q_i$ give 4 and 2 crossings. For $\ar=3\times10^{-5}$, the four counts are 10, 12, 10, and 4, whereas the baseline has 8. For $\ar=10^{-4}$, every count remains 5, but the redshifts of the crossings change. Since changing $U_i$ also resets the potential normalization, the comparison belongs to this specified family of initial states. Within that family, $\alpha$ and the GR reference parameters alone do not fix the crossings.

To separate curvature-data sensitivity from a scalar-potential adjustment, we also keep $(\varphi_i,v_i,U_i)$ fixed within each coupling row of Table~\ref{tab:initialvalues}. We replace either $r_i$ or $q_i$ by $(1\pm0.01)$ times its baseline value and solve $\mathcal C_i=0$ for the positive $E_i$ root connected to the baseline, within the bracket $[0.8E_{i,\mathrm{base}},1.2E_{i,\mathrm{base}}]$. Matter, radiation, the vector normalization and the potential normalization are unchanged. The initial Hubble rate is therefore an output of this constraint solve, not an additional fixed datum. This is a comparison of constrained gravitational initial states at fixed scalar initial data.

\begin{table}[th]
\centering\small
\caption{Crossing counts for fixed initial scalar data. We obtain $E_i$ from the constraint after multiplying the indicated baseline curvature variable by $1\pm0.01$, as denoted by the column signs.}\label{tab:fixedscalar}
\begin{tabular}{@{}cccccc@{}}\hline\hline
$\ar$&base&$r_i-$&$r_i+$&$q_i-$&$q_i+$\\\hline
$10^{-5}$&2&16&14&2&4\\
$3\times10^{-5}$&8&10&12&4&10\\
$10^{-4}$&5&5&5&5&5\\
\hline\end{tabular}
\end{table}
The new comparison confirms that changes in crossing multiplicity do not require readjusting the scalar potential. The initial Hubble shifts are below $1.80\times10^{-4}$ in magnitude. Radau and BDF give identical crossing counts, with maximum relative Hubble difference below $1.6\times10^{-10}$ and maximum crossing-redshift difference below $6.6\times10^{-8}$. Every crossing count is unchanged when the root-search grid is increased from 10001 to 20001 points. The sampled effective density remains positive in all 15 baseline and perturbed histories, with a minimum above $0.845$ in the units of Eq.~\eqref{eq:effrho}. No probability distribution for the initial states follows from these local comparisons.

\subsection{Early-time reference diagnostics}\label{sec:calibration}
On the GR reference background we obtain
\begin{equation}
r_{\mathrm{GR}}(1100)=1.248307893\times10^9.
\end{equation}
Evaluating $F=1+2\ar r$ on that reference gives approximately $2.497\times10^4$, $7.490\times10^4$, and $2.497\times10^5$ for the three trial couplings. These are evaluations on the GR reference, used to check the assumption of a small departure from GR. They do not follow the quadratic evolution through recombination.
As an illustrative reference criterion, imposing $|F-1|<0.01$ at $z=1100$ gives
\begin{equation}
\ar<4.0054\times10^{-12}.
\end{equation}
The 1\% criterion serves only to gauge the reference expansion; it is not a constraint inferred from observations. In particular, the quadratic examples cannot consistently inherit the GR sound horizon or a compressed CMB likelihood without computing their early evolution. Small $\ar$ does not settle the scalaron initial-value problem: its rapid mode remains stiff, and both the selected branch and its excitation must be stated.

\section{GR-connected response}\label{sec:response}

\begin{figure}[t]
\centering
\includegraphics[width=0.96\textwidth]{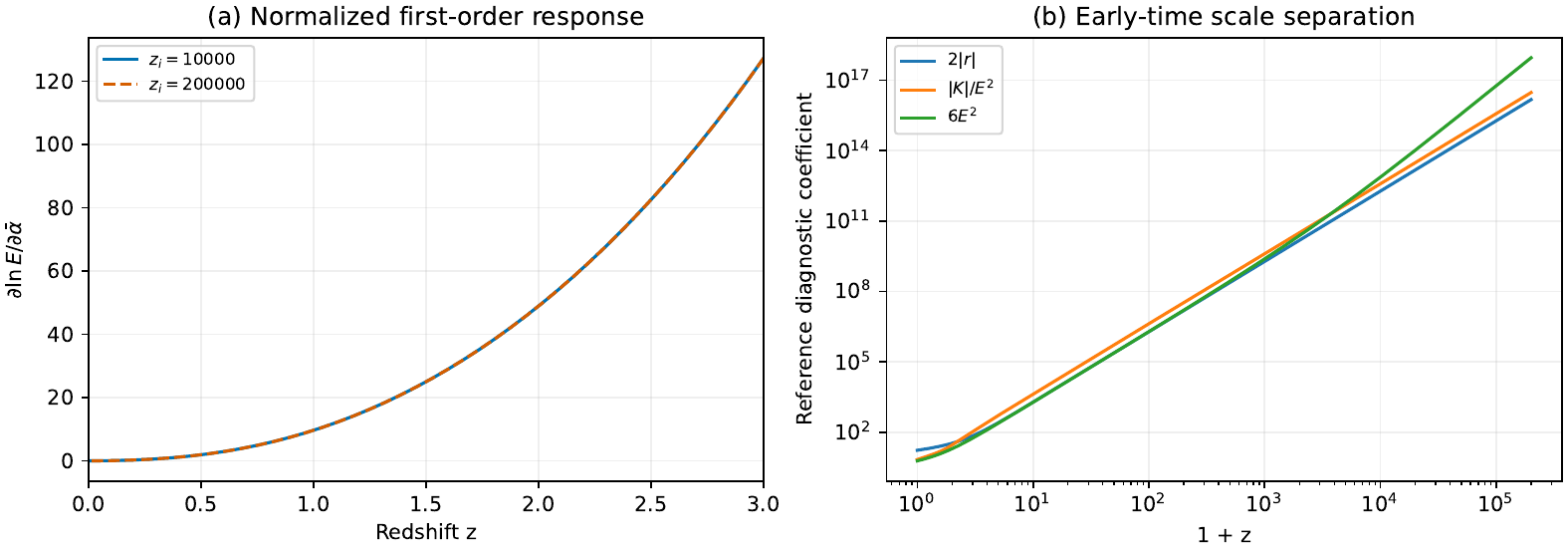}
\caption{(a) First-order response with $e_1(0)=0$. Matching at $10^4$ and $2\times10^5$ yields nearly coincident curves. (b) The coefficients for all three reference checks become larger at earlier times. These panels show the response and calibration tests, rather than inferred parameter distributions.}\label{fig:response}
\end{figure}

We next calculate the response obtained by perturbing the GR background while setting aside the independent homogeneous scalaron oscillation. Figure~\ref{fig:response} displays the resulting response and the reference tests. Suppressing that free mode is part of the order-reduction prescription~\cite{Simon1990,ParkerSimon1993,Barros}, not a statement about the initial conditions of every solution in the full theory. Writing $\mathcal S=(\rho-3P)/\Mp^2$, the quadratic trace equation gives
\begin{equation}
R=\mathcal S-6\alpha(\ddot{\mathcal S}+3H\dot{\mathcal S})+O(\alpha^2).
\end{equation}
This expansion requires source evolution on timescales long compared with $m_R^{-1}$, together with initial data that do not excite the free oscillation. The numerical benchmarks illustrate what changes when the latter condition is relaxed.
For completeness, the quadratic action is equivalent, for $\alpha\ne0$, to an auxiliary curvature field $\Phi$ with gravitational Lagrangian
\begin{equation}
\frac{\Mp^2}{2}\left[\Phi R-\frac{(\Phi-1)^2}{4\alpha}\right],\qquad \Phi=1+2\alpha R.
\end{equation}
For $\Phi>0$ its Einstein-frame scalar coordinate is $\sigma=\sqrt{3/2}\,\Mp\ln\Phi$. The corresponding curvature potential is $\Mp^2(\Phi-1)^2/(8\alpha\Phi^2)$, as follows from the standard metric $f(R)$ scalar--tensor representation~\cite{DeFelice}. The transformed SVT sectors interact with this scalar; the curvature potential alone does not establish their coupled stability.
\subsection{Normalized tangent equations}
All reference quantities below are evaluated along the GR solution. At fixed matter densities, vector amplitude, $s$, and $\lambda$, set $E=E_g+\ar e_1+O(\ar^2)$, $v=v_g+\ar v_1+O(\ar^2)$, and $U=U_g+\ar U_1+O(\ar^2)$. Define
\begin{align}
K_g&=\frac{r_g^2}{6}-2E_gq_g-2r_gE_g^2,\\
D_g&=2E_g+\frac{2sd_{\chi,g}}{E_g}.
\end{align}
Linearizing the Friedmann constraint yields
\begin{equation}
e_1=\frac{v_gv_1/3+U_1+K_g}{D_g}.
\end{equation}
The scalar equations then form a nonstiff linear response system:
\begin{align}
(v_1)_N&=-3v_1+3\lambda\left(\frac{U_1}{E_g}-\frac{U_ge_1}{E_g^2}\right),\\
(U_1)_N&=-\lambda\left[\frac{v_gU_1+U_gv_1}{E_g}-\frac{U_gv_ge_1}{E_g^2}\right].
\end{align}
These are derivatives of the order-reduced background at $\ar=0$, rather than an integration of a nonlinear quadratic cosmology with finite coupling. $r_g$ and $q_g$ are obtained analytically from the conserved GR trace and its derivative, avoiding cancellation in radiation domination.
We use the leading radiation-era scalar matching relation $v_i\simeq3\lambda U_i/(5E_i)$ to specify its linearized initial variation. This scalar matching is itself approximate. One forced response and one homogeneous scalar-amplitude response are integrated. Their linear combination is chosen to impose
\begin{equation}
e_1(0)=0.
\end{equation}
Thus the present-day Hubble normalization is fixed consistently at first order. The vector amplitude is held fixed: its fraction of the effective total DE density need not remain the original $f_\chi$ after adding curvature terms.

\subsection{Reference criteria for scale separation}
Throughout the matching interval, we impose three illustrative tolerances on the reference solution:
\begin{equation}
2\ar|r_g|\le0.01,\quad \ar|K_g|/E_g^2\le0.01,\quad 6\ar E_g^2\le0.01.
\end{equation}
The last condition is $H_g^2/m_R^2\le0.01$. The first probes $F-1$, the second the explicit gravitational forcing, and the third the source-timescale hierarchy. The tolerances are chosen for these diagnostics; they supply neither an observationally inferred prior nor a local-gravity bound, and they do not complete the CMB calibration. For positive $\ar$, their common reference threshold is
\begin{equation}
\ar_{\mathrm{diag}}=\min_N\left\{\frac{0.01}{2|r_g|},\frac{0.01E_g^2}{|K_g|},\frac{0.01}{6E_g^2}\right\}.
\end{equation}
For matching from $z_i=10^4$ and $2\times10^5$, the thresholds are respectively $1.342\times10^{-15}$ and $1.102\times10^{-20}$. Their dependence on matching epoch is essential; they must not be presented as a universal bound on the theory.
The maximum normalized first-order response on $0\le z\le3$ is approximately
\begin{equation}
\max\left|\frac{e_1}{E_g}\right|=127.36
\end{equation}
for both matching epochs. Multiplication by the corresponding diagnostic coupling gives formal first-order changes of approximately $1.71\times10^{-13}$ and $1.40\times10^{-18}$ in $H/H_g$. We obtain these small changes from the asymptotic response, rather than by resolving them in a full nonlinear integration. Their higher-order errors have not been bounded.

\subsection{Response validation and interpretation}
The response is stable when the integration tolerance is tightened and the maximum step is halved. An additional finite-parameter test uses an algebraically constrained order-reduced model over $0\le z\le3$, with the same differentiated scalar matching family and a re-shot $E(0)=1$. Its relative difference from the tangent solution is $1.63\times10^{-4}$ at a test parameter $10^{-6}$ and $8.14\times10^{-5}$ at $5\times10^{-7}$, consistent with first-order truncation. This comparison checks the response implementation but leaves the nonlinear early-time scalaron branch untested.
At the reference point, the adiabatic quadratic response connected perturbatively to GR gives little sensitivity in late-time expansion contours when the stated early-time criteria are imposed. The pronounced crossing changes in the benchmarks occur in another regime, where the additional initial data matter. A general exclusion theorem would require control of nonlinear branches, excited scalaron histories, broader SVT parameters, and perturbations. None is supplied by this first-order result.

Exact finite-coupling checks on $0\le z\le3$ and an early radiation window are given in Appendix~\ref{app:validation}. They test specified initial-value problems and the local expansion, while the first-order response above remains a formal branch construction.
\section{Vector-dependent fast mode}\label{sec:fast}
The curvature-sector mass coefficient $1/(6\alpha)$ does not by itself describe the frequency of a rapidly varying homogeneous mode after the SVT response is included. The vector branch obeys $\rho_\chi\propto H^{-2s}$, with
\begin{equation}
\rho_\chi-3P_\chi=4\rho_\chi-2s\rho_\chi\epsilon.
\end{equation}
For a homogeneous perturbation, $\delta\rho_\chi=-2s\rho_\chi\delta H/H$ and $\delta R=6\delta\dot H+24H\delta H$. Hence
\begin{align}
\frac{\delta(\rho_\chi-3P_\chi)}{\Mp^2}
={}&-s\Omega_\chi\delta R\nonumber\\
&+\frac{4s(s+1)\epsilon\rho_\chi}{\Mp^2}\frac{\delta H}{H},
\end{align}
where $\Omega_\chi=\rho_\chi/(3\Mp^2H^2)$.
If the perturbation frequency is much greater than $H$, the second term is subleading because $\delta H\simeq\delta R/(6i\omega)$. Dust, radiation, and canonical scalar responses are also subleading at this frequency when the background is slowly varying. The quadratic trace then has leading homogeneous frequency
\begin{equation}
\omega_{\mathrm{fast}}^2\simeq\frac{1+s\Omega_\chi}{6\alpha}.
\end{equation}
For smooth general metric $f(R)$ coefficients, the same principal balance gives
\begin{equation}
\omega_{\mathrm{fast}}^2\simeq\frac{F-Rf_{,RR}+s\Omega_\chi}{3f_{,RR}},
\end{equation}
provided background-coefficient variations and the neglected terms remain slow compared with this mode. This is a homogeneous short-timescale statement, not a finite-wavelength dispersion relation or a full kinetic-matrix result.
At the Table~II reference today, $s\Omega_{\chi,0}=0.11234$, so the quadratic fast frequency is about $1.0547$ times the isolated curvature-sector value. We validate the leading coefficient against the complex eigenvalues of the frozen homogeneous Jacobian at $z=0,1,3$ and four couplings from $10^{-5}$ to $10^{-8}$. At $10^{-8}$, relative frequency errors are $5.14\times10^{-8}$, $3.03\times10^{-7}$, and $2.28\times10^{-6}$, respectively; they decrease as the coupling decreases.
For positive $\alpha$ and $s\Omega_\chi>-1$, this leading fast frequency is real. That is one local necessary diagnostic. Ghost freedom, gradient stability, finite-frequency modes, and global nonlinear evolution still require the coupled perturbation calculation. The SVT analyses in Refs.~\cite{SVTCosmology2018,SVTDarkEnergy2018,GaugeReady2018} provide useful starting points, but their perturbation matrices must be extended to include the independent scalaron.
\subsection{Damping and amplitude transport}
Let $\Omega_\chi=\rho_\chi/(3\Mp^2H^2)$ and $m^2=(6\alpha)^{-1}$. In the high-frequency homogeneous limit, the vector trace perturbation has the form
\begin{equation}
\delta\mathcal S_\chi=-s\Omega_\chi\delta R+B\delta H/H,
\end{equation}
where $B=4s(s+1)\epsilon\rho_\chi/\Mp^2$. The exact geometric relation $\delta\dot H+4H\delta H=\delta R/6$ implies, to leading oscillatory order, $\delta H/H\simeq-\delta\dot R/(6H\omega^2)$. The principal curvature equation therefore supplies
\begin{align}
\omega^2&\simeq m^2(1+s\Omega_\chi),\\
\frac{\Gamma}{H}&\simeq3+\frac{2s(s+1)\epsilon\Omega_\chi}{1+s\Omega_\chi},
\end{align}
with $\delta\ddot R+\Gamma\delta\dot R+\omega^2\delta R\simeq0$. Other source responses and background-derivative terms enter subleading corrections in this fast limit. This is not a finite-wavelength dispersion relation.
For slowly varying coefficients the leading WKB solution is $\delta R=A_R\cos\Theta$, $\dot\Theta=\omega$, and $A_R\propto\omega^{-1/2}\exp[-\tfrac12\int\Gamma\,dt]$. Since $(\Omega_\chi)_N=-2(s+1)\epsilon\Omega_\chi$ on the temporal-vector branch, this gives
\begin{equation}
\frac{A_R(N)}{A_R(N_a)}\simeq e^{-3(N-N_a)/2}
\left[\frac{1+s\Omega_\chi(N)}{1+s\Omega_\chi(N_a)}\right]^{1/4}.
\end{equation}
\label{para:wkbtransport}
The corresponding relative Hubble envelope is $A_R/(6H\omega)$. For the frozen Jacobian at $z=0,1,3$, decreasing $\alpha$ brings the real parts of the fast eigenvalues towards $-\Gamma/(2H)$. This is the leading damping predicted by the asymptotic calculation. The comparison checks that local coefficient, leaving the full stability problem open.

\subsection{Initial-data restriction on the scalaron}
The absence of a free scalaron oscillation is a choice of initial data. It does not follow merely from taking $\alpha$ small. An excited scalaron may have an independently specified energy density, as in the dark-matter scenarios studied elsewhere~\cite{Shtanov}. To assess such a history, one must prescribe its amplitude and phase and follow its early-time evolution with a suitable calibration. Our result for negligible corrections on the unexcited branch says nothing about those cases. We therefore do not exclude excited quadratic branches in general, nor do we use the three selected reference diagnostics to assign a prior to them.

\section{Conditional distance sensitivity}\label{sec:distance}
The oscillatory ansatz below uses the amplitudes and mean-feedback prescriptions specified in Appendix~\ref{app:validation}. Its distance bounds are conditional on those approximations and do not bound an unknown global background error.

We assess the observable sensitivity of the transported rapid-mode ansatz at the Table~II marginal means. This is an approximation diagnostic, not a likelihood fit or a bound on $\ar$. For broader cosmological parameter context, see the Planck 2018 analysis~\cite{Planck2018}. We use the bundled DESY5 supernova data~\cite{DESY5} and DESI DR2 BAO data~\cite{DESIDR2}: 1829 supernova distances and 13 correlated BAO entries. The supernova covariance is the supplied systematic covariance plus the squared tabulated statistical errors. Its constant absolute-magnitude direction is projected out, following the Cobaya~\cite{Cobaya} DESy5 implementation in release 3.6.2.
For flat geometry, let $\chi(z)=\int_0^z dz'/E(z')$. Then $D_M=c\chi/H_{\mathrm{ref}}$, $D_H=c/(H_{\mathrm{ref}}E)$, and $D_V=(zD_M^2D_H)^{1/3}$. The supernova luminosity distance uses $(1+z_{\mathrm{HEL}})D_M(z_{\mathrm{HD}})$. All BAO comparisons hold the common sound horizon at $147.64749\,\mathrm{Mpc}$ solely as a scaling reference. This value should not be interpreted as a prediction of the quadratic model: we have not established the recombination calibration or the validity of the compressed-CMB likelihood in this setting.

Let $h=A_R/(6H\omega)$ be the fractional envelope of the Hubble oscillation and let $m$ denote the frozen mean estimate in Sec.~\ref{sec:transport}. We set $b=h+|m|$ and introduce the dimensionless quantities $w=e^{-N}/E$ and $g=w h/(\omega/H)$. After integration by parts, the leading oscillatory distance correction is bounded by the values of $g$ at the endpoints and its total variation between them. On the tested interval, its logarithmic slope can be written analytically as
\begin{equation}
\frac{g_N}{g}=-\frac52-\epsilon+
\frac{3s(s+1)\epsilon\Omega_\chi}{2(1+s\Omega_\chi)}<0.
\end{equation}
Thus the bound reduces to $2g(N_z)$, avoiding direct integration of millions of oscillations. Adding a bound on the reciprocal remainder gives, within this ansatz,
\begin{equation}
\frac{|\delta\chi|}{\chi}\le\frac{2g(N_z)+
\int_{N_z}^0w\left[|m|+\frac{b^2}{1-b}\right]dN}{\chi}.
\end{equation}
The radial-distance bound is $|\delta D_H|/D_H\le b/(1-b)$. We apply exact logarithmic and product intervals when converting these envelopes to supernova magnitudes and $D_V$. The radial bound is pointwise; we do not model additional cancellation across a BAO redshift bin.
\begin{table}[t]
\centering
\caption{Conditional ansatz envelopes with the frozen mean estimate at the reference parameters. These are neither exact-model errors nor measured coupling limits.}\label{tab:observables}
\begin{tabular}{@{}ccc@{}}
\hline\hline
$\ar$&$10^{-15}$&$10^{-16}$\\
\hline
max $|\delta\mu|$ (mag)&$7.1\times10^{-8}$&$2.42\times10^{-12}$\\
max $|\delta B_i|/\sigma_i$&$1.63\times10^{-2}$&$6.41\times10^{-5}$\\
SN covariance norm&$2.58\times10^{-5}$&$6.11\times10^{-10}$\\
BAO covariance norm&$3.08\times10^{-2}$&$1.21\times10^{-4}$\\
\hline
\end{tabular}
\end{table}
We account for correlations with a conservative bound on the covariance norm. If the componentwise shift is bounded by $a_i$, then for the appropriate inverse covariance $W$ we have $\delta^TW\delta\le a^T|W|a$. Here the absolute value is taken separately for each matrix entry. Table~\ref{tab:observables} gives the square root of this upper bound, with $W$ including the magnitude projection for supernovae. For $\ar=10^{-15}$ at the reference parameters, this gives conditional $|\Delta\chi^2|$ bounds of 0.00209 for supernovae and 0.1864 for BAO, evaluated as $2\sqrt{\chi^2 Q}+Q$ with $Q=a^T|W|a$. The quoted values change with the residual prescription, the fixed sound-horizon scale, and the assumed oscillatory form. For the branch without free scalaron excitation, the formal first-order supernova shifts are $6.83\times10^{-15}$ and $6.83\times10^{-16}$ magnitudes. We evaluate these tiny responses from the distance derivatives, since subtracting the two nearly identical distances would lose numerical precision.

On refining the distance grid from 10001 to 20001 points, the diagnostic envelopes and formal shifts change by less than $1.4\times10^{-11}$ relatively. The observable integrals are consequently insensitive to this particular grid refinement. The refinement does not determine the global background error or verify either continuous nonlinear backreaction or accuracy over a parameter posterior.
If we instead include the integrated mean feedback and impose endpoint normalization, the conditional maximum supernova modulus envelopes become $5.74\times10^{-8}$ and $2.20\times10^{-12}$ magnitudes. The maximum BAO envelopes remain $0.01632\sigma_i$ and $6.409\times10^{-5}\sigma_i$. The corresponding supernova covariance-norm bounds are $1.864\times10^{-5}$ and $5.021\times10^{-10}$, while the BAO values remain $0.03080$ and $1.209\times10^{-4}$. This extends the sensitivity diagnostic to the constructed averaged family; the unknown exact global error and sound-horizon calibration are still outside these bounds.

\subsection{Formal coupling information}\label{sec:information}
Successful numerical checks do not ensure that a coupling is measurable. We assess the local information in the formal, normalized, unexcited branch separately from the families obtained by matching at finite order, which retain residual rapid modes. We write its observable derivative as $J=\partial\boldsymbol\mu/\partial\ar$, after projecting out the constant supernova magnitude direction. With the reference cosmological parameters and the common sound horizon held fixed, the covariance-weighted information is
\begin{equation}
I_{\ar}=J_{\mathrm{SN}}^{T}W_{\mathrm{SN}}J_{\mathrm{SN}}
+J_{\mathrm{BAO}}^{T}W_{\mathrm{BAO}}J_{\mathrm{BAO}}.
\end{equation}
We integrate the analytic first-order derivatives directly, avoiding the subtraction of nearly identical distances. At the reference parameters, this gives $I_{\ar}^{-1/2}=1.339\times10^{-4}$. This scale describes the local derivative response, but lies far beyond the range in which the expansion is controlled. It therefore cannot be interpreted as a measured uncertainty or an observational bound on $\ar$.

We also project out the BAO sound-horizon calibration direction, $u_i=D_i/r_d$, using
\begin{equation}
W_{\mathrm{BAO}}^{\perp}=W_{\mathrm{BAO}}
-\frac{(W_{\mathrm{BAO}}u)(W_{\mathrm{BAO}}u)^T}
{u^TW_{\mathrm{BAO}}u}.
\end{equation}
The resulting combined inverse-information scale is $1.592\times10^{-4}$. This is a local nuisance-direction projection, not a calibrated quadratic sound horizon or a marginal posterior. Allowing additional nuisance parameters cannot increase this local Fisher information without additional information or priors.
For the chosen reference diagnostic interval $0\le\ar\le\ar_{\mathrm{diag}}$, the formal data displacement has norm $\ar_{\mathrm{diag}}\sqrt{I_{\ar}}$. With matching at $z_i=10^4$, this norm is $1.002\times10^{-11}$; matching at $z_i=2\times10^5$ gives $8.229\times10^{-17}$ because its reference diagnostic interval is much smaller. These intervals use the previously chosen 1\% tolerances, not a measured constraint. At the reference point the absolute change in the fixed-calibration quadratic Gaussian $\chi^2$, using the linear observable response throughout the chosen interval, is bounded by $9.96\times10^{-12}$ for the lower matching redshift.
\begin{table}[t]
\centering
\caption{Formal unexcited-branch information diagnostics. The scale $I_{\ar}^{-1/2}$ is not a measured uncertainty; the diagnostic coupling interval is chosen from reference tolerances.}\label{tab:information}
\begin{tabular}{@{}ccc@{}}
\hline\hline
$z_i$&$\ar_{\mathrm{diag}}$&$\ar_{\mathrm{diag}}\sqrt{I_{\ar}}$\\
\hline
$10^{4}$&$1.342\times10^{-15}$&$1.002\times10^{-11}$\\
$2\times10^{5}$&$1.102\times10^{-20}$&$8.229\times10^{-17}$\\
\hline
\end{tabular}
\end{table}
Ten illustrative tests change $H_{\mathrm{ref}},\Omega_m,f_\chi,s_\chi$, or $\lambda$ individually by $\pm10\%$ and repeat the reference and derivative calculation at $z_i=10^4$. Their displacement norms range from $9.35\times10^{-12}$ to $1.06\times10^{-11}$. These are prescribed robustness tests, not coverage of a joint prior or posterior. They support local observational degeneracy of this formal branch within the tested domain. They do not exclude larger-coupling nonlinear branches or independently excited scalaron histories. The residual-mode covariance envelopes above concern a different initial-data family and must not be combined with this unexcited derivative as one posterior.
Consequently, these calculations cannot supply a meaningful measured $\ar$ contour for the restricted formal branch. Longer sampling cannot compensate for this lack of local information. Other quadratic histories would have to be analysed with a stated distribution of scalaron initial data, reliable evolution over the full history, and an early-time calibration consistent with that evolution. No observational posterior is claimed here.

\section{Interpretation and limits}\label{sec:discussion}
The field-equation identities and the behaviour of individual trajectories have different scopes. The identity $C_N=-4C$ follows from the background equations, but the effective phantom-crossing count is set by the constrained initial state. The small distance shifts belong to the normalized expansion with no free scalaron excitation and cannot be attributed to all solutions of the quadratic theory. The illustrative coupling range is likewise chosen through reference tolerances, from which no observational upper bound on $\ar$ follows.

The first crossing comparison changes the curvature data together with the potential required by the constraint. In Table~\ref{tab:fixedscalar}, we instead hold the scalar data fixed and solve the constraint for the initial Hubble rate. Curvature-data dependence remains in this second comparison. Neither test establishes a likelihood preference for phantom behavior. Our fast-frequency result concerns the temporal vector coupled to a homogeneous, rapidly varying perturbation; kinetic and gradient stability at finite wavelengths still have to be calculated.

Running MCMC for longer cannot add the local information about $\ar$ missing from this restricted regular response. The derivative test in Sec.~\ref{sec:information} concerns only that response on the stated interval. Other quadratic histories may behave differently. To analyse excited histories observationally, one must prescribe the initial amplitude and phase, evolve them reliably to the observation epoch, and calculate the acoustic scale from their early-time dynamics.
\section{Conclusions}\label{sec:conclusions}
In the quadratic SVT extension, the background equations form a constrained evolution system with Friedmann residual $C_N=-4C$. Along the evolution, the acceleration residual is proportional to the Friedmann constraint. We therefore rely on independent numerical integrations to check the chosen solutions, since comparing the two residuals alone supplies no second validation.

Both the crossing count and the crossing redshifts respond to changes in the initial curvature data in our late-time benchmarks, including the tests with fixed scalar data. We also find that the temporal-vector background enters the leading frequency and damping of the homogeneous scalaron. These are results for background evolution and its homogeneous high-frequency limit. Whether the coupled finite-wavelength perturbations avoid ghost and gradient instabilities remains to be established.

For the normalized order-reduced solution, the calculated distance shift is extremely small throughout the selected reference interval. This leaves little local sensitivity to $\ar$ in the restricted family and provides no measured constraint on the coupling. Our checks of local integrations, reduced evolution, and transported oscillations address separate constructions; they have yet to establish one continuous exact history. Before using the model for inference, the nonlinear equations must be followed from the radiation era with accumulated errors under control and a sound horizon computed for the same evolution.

In the future work, we will aim to evolve the original quadratic field equations continuously from the radiation era to late times, with controlled scalaron initial conditions and a quantitative estimate of accumulated numerical errors. A separate analysis of the scalar, vector, and tensor perturbations is needed to determine when ghost and gradient instabilities are absent. We must also test the local gravitational constraints and compute the sound horizon within the modified expansion history. Only then would a comparison with supernova distances, baryon acoustic oscillations, the cosmic microwave background, and structure growth have the required dynamical basis. It could determine whether the curvature evolution and effective phantom-divide crossings can be distinguished observationally from those of the corresponding general-relativistic SVT background.


\section*{Acknowledgments}
The work of Kazuharu Bamba was supported in part by the JSPS KAKENHI Grant No.~25KF0176 and a grant-in-aid of academic research of the Yamaguchi Scholarship Foundation.

\FloatBarrier
\appendix
\section{Numerical validation and conditional mode tests}\label{app:validation}
\suppressfloats[t]

The tests below cover three types of calculation: direct integration of the quadratic equations, order reduction, and averaging. We identify the starting data for each test and quote residuals only where they have been evaluated.

Table~\ref{tab:scope} separates the calculation families. For each test below, we distinguish comparisons against the original field equations from checks performed entirely inside an approximate model.

\subsection{Local check against the nonlinear equations}\label{sec:local}
To check the expansion locally, we solve the original quadratic equations from $z=3$ to $z=0$. We use the first-order response and its derivatives to set the initial $H$, scalar velocity, potential, curvature, and curvature velocity. All derivatives at the starting slice are calculated with one-sided stencils lying inside the computed interval. A correction to $U_i$ of order $O(\ar^2)$ then removes the residual violation of the initial constraint.

For every finite coupling, we vary the scalar-amplitude response parameter and repeat the full integration until $E(0)=1$. This defines the initial-data family for the local comparison. It does not determine a unique extension of that family into the radiation era.
\begin{table}[th]
\centering
\caption{Comparison with the full nonlinear response. We use $\eta_H$ for the maximum relative difference between the nonlinear finite-difference and first-order responses, normalized by the maximum first-order response.}\label{tab:nonlinear}
\begin{tabular}{@{}ccc@{}}
\hline\hline
$\ar$&$\eta_H$&$\max|H_{\mathrm R}-H_{\mathrm B}|/H_{\mathrm R}$\\
\hline
$10^{-6}$&$1.347\times10^{-3}$&$6.79\times10^{-12}$\\
$3\times10^{-7}$&$3.975\times10^{-4}$&$2.56\times10^{-12}$\\
$10^{-7}$&$1.325\times10^{-4}$&$7.92\times10^{-13}$\\
\hline
\end{tabular}
\end{table}
We use $\mathrm R$ for Radau and $\mathrm B$ for BDF. All three cases give $|E(0)-1|$ below $4\times10^{-15}$ and a scaled Friedmann residual below $10^{-12}$. Reducing the initial derivative-stencil step by half changes $H$ by at most $2.1\times10^{-13}$ relatively. The response error falls with the coupling, as a first-order expansion should. We thus find local nonlinear agreement with the imposed normalization, without determining the global early-time branch.
\subsection{Exact radiation window}\label{sec:radiation}
For the early-time test, we solve the original quadratic equations on $10^3\leq z\leq10^4$ with the prescribed first-order curvature-matched data. The initial scalar potential is kept fixed, while $H_i$ is found from the exact Friedmann constraint. Matching at this order can leave a homogeneous scalaron contribution. As the solution evolves across the interval, the radiation fraction decreases from about $0.75$ to $0.23$.

For $\bar\alpha=\alpha H_{\mathrm{ref}}^2=10^{-15},3\times10^{-16},10^{-16}$, the maximum fractional deviations of $H$ from the GR reference are $1.703\times10^{-3}$, $5.586\times10^{-4}$, and $1.907\times10^{-4}$, respectively. The corresponding maximum discrepancies from the first-order response, divided by the maximum magnitude of that response, are $0.1176$, $0.03528$, and $0.01176$. Radau and BDF agree in $H$ to within $2.2\times10^{-10}$ in relative terms, while the scaled Friedmann residual stays below $2.6\times10^{-12}$. With half the initial derivative-stencil step, the change in $H$ is below $6.1\times10^{-11}$. We also find $F$ positive over the whole interval.

These early-time runs reproduce the numerical solution and tend towards the first-order response at smaller $\bar\alpha$. They do not yet give a normalized exact history from the radiation era to today. Nor do they establish full perturbative stability or constrain $\alpha$ observationally.
\subsection{Order-reduced background evolution}\label{sec:slow}
A frequency count based on the isolated curvature mass and the GR expansion yields roughly $1.94\times10^6$ cycles from $z=10^4$ to today. Direct time stepping must therefore resolve an enormous number of periods. This count is a computational guide based on the reference background, rather than a frequency integral evaluated along an exact quadratic-model history.

For the finite-coupling comparison we instead evolve an order-reduced closure. At each step, the current scalar variables $(v,U)$ and a trial expansion rate $E$ determine $\epsilon_g$, $r_g$, and $q_g$ through the algebraic GR source relations. We define
\begin{equation}
K_g=r_g^2/6-2Eq_g-2r_gE^2.
\end{equation}
At each integration point solve
\begin{equation}
E^2=d_\mathrm{m}+d_\mathrm{r}+d_\chi+v^2/6+U+\ar K_g
\end{equation}
and integrate the scalar equations. Integrating this algebraic closure retains nonlinear dependence generated by its first-order forcing, but does not restore the higher-order terms discarded in its derivation or the freely oscillating curvature degree of freedom. The radiation-matching prescription sets the initial scalar velocity, and shooting in $U_i$ enforces $E(0)=1$. At $\ar=10^{-15},3\times10^{-16},10^{-16}$, the reduced-system solutions from DOP853 and Radau differ in $H$ by less than $4.0\times10^{-14}$, and the endpoint normalization holds to $3\times10^{-16}$.

We recover the geometric curvature by differentiating the algebraic constraint implicitly. The residual obtained when this curvature is inserted into the original trace equation is
\begin{equation}
D=r-\mathcal T+6\ar E^2[r_{NN}+(3+\epsilon)r_N].
\end{equation}
\begin{table}[t]
\centering
\caption{Original-equation defects of the reduced solution on $N_i+0.008\leq N\leq-0.008$. The normalization is $\max(|\mathcal T|,1)$ for the trace defect and $E^2$ for the Friedmann residual.}\label{tab:slow-defects}
\begin{tabular}{@{}ccc@{}}
\hline\hline
$\ar$&$\max|D|/\max(|\mathcal T|,1)$&$\max|\mathcal C|/E^2$\\
\hline
$10^{-15}$&$1.167\times10^{-2}$&$4.449\times10^{-4}$\\
$3\times10^{-16}$&$1.035\times10^{-3}$&$3.973\times10^{-5}$\\
$10^{-16}$&$1.145\times10^{-4}$&$4.405\times10^{-6}$\\
\hline
\end{tabular}
\end{table}
Halving the differentiation stencil from $0.002$ to $0.001$, on the same sampling grid, barely changes the largest defects. This checks the evaluation of those residuals; it cannot be used as a bound on the accumulated error in the reduced trajectory.

For the comparison on $10^3\leq z\leq10^4$, both evolutions start with the same scalar potential and velocity. The largest relative differences in $H$ are $2.3194\times10^{-4}$ at $\ar=10^{-15}$ and $2.3227\times10^{-6}$ at $\ar=10^{-16}$. The quoted differences include the initial offset in $H$ required by the two closures. At the final redshift $z=10^3$, they are $4.0121\times10^{-5}$ and $5.0843\times10^{-7}$, respectively.

Agreement over the early window tests the closure where the two calculations overlap. It leaves open the late-time error and whether finite-order matching has eliminated the homogeneous mode. The normalized reduced trajectory should therefore not be identified with a verified solution of the original equations spanning radiation domination to the present, nor used as a fitted cosmological history.

\subsection{Transport of the residual oscillation}\label{sec:transport}
The free homogeneous oscillation omitted in formal order reduction may survive when the initial data are matched only to finite order. We determine the residual oscillation by fitting it, rather than assuming that the matched data define an exact unexcited branch.

The frequency, damping, and leading WKB amplitude used here are derived in Sec.~\ref{sec:fast}, with relative Hubble envelope $A_R/(6H\omega)$. Over $1000\leq z\leq3000$, we fit the difference between the curvature obtained from the original equations and that of the reduced scalar family. We integrate the phase using $H$ from the full nonlinear solution on this interval and include a smooth nuisance polynomial to account for nonoscillatory differences between the closures. Using polynomial degrees 3, 5, and 7 changes the fitted mode amplitude by less than $6.0\times10^{-7}$ in relative terms at $\ar=10^{-15}$ and $1.3\times10^{-8}$ at $\ar=10^{-16}$. The curvature-fit residual is below $0.09\%$ of the oscillation RMS.

Independent fits to $H$ confirm the curvature-to-Hubble amplitude relation to within $0.0025\%$ and $0.00034\%$, respectively. Their waveform residuals are $0.207\%$ and $0.0665\%$ of the Hubble oscillation RMS.
\begin{table}[t]
\centering
\caption{Present-day envelopes from leading WKB transport of the specified residual mode on the GR reference. The values are asymptotic estimates and carry neither an exact-evolution nor an observational-constraint interpretation.}\label{tab:transport}
\begin{tabular}{@{}ccc@{}}
\hline\hline
$\ar$&$A_R(0)/H_{\mathrm{ref}}^2$&$|\delta H|_{\mathrm{env}}/H_{\mathrm{ref}}$\\
\hline
$10^{-15}$&$8.776\times10^{3}$&$1.074\times10^{-4}$\\
$10^{-16}$&$1.09\times10^{2}$&$4.219\times10^{-7}$\\
\hline
\end{tabular}
\end{table}
These residual-mode envelopes describe the initial-data family matched at finite order. They should be considered separately from the negligible formal late-time response of the strictly unexcited order-reduced branch. A large curvature amplitude can accompany a small Hubble amplitude because $\omega\gg H$. We have not established the exact late-time phase, the nonlinear averaged backreaction, error bounds over the full interval, or a prescription selecting a unique branch with no free mode at any order. Neither the smooth reduced solution nor the WKB diagnostic alone therefore provides a sufficient basis for a validated observational posterior.

\subsubsection{Leading phase-average contribution}
The instantaneous oscillation envelope and its mean contribution describe different effects. Holding the slow background fixed over one period, we write $\delta R=A_R\cos\Theta$, $\delta H=A_R\sin\Theta/(6\omega)$, and $\delta\dot R=-\omega A_R\sin\Theta$. Averaging then gives $\langle\delta H^2\rangle=A_R^2/(72\omega^2)$. Expanding the original Friedmann constraint and the vector density to quadratic order yields an effective source offset
\begin{equation}
B_{\mathrm{osc}}\simeq\frac{\alpha A_R^2[1+s(s+2)\Omega_\chi]}{6(1+s\Omega_\chi)}.
\end{equation}
At fixed scalar data, the corresponding frozen mean-Hubble shift is
\begin{equation}
\frac{\delta\bar H}{H}\simeq\frac{\alpha A_R^2[1+s(s+2)\Omega_\chi]}{12H^2(1+s\Omega_\chi)^2}.
\end{equation}
The quadratic scalaron Einstein-frame energy provides a separate diagnostic,
\begin{equation}
\frac{\langle\rho_\sigma\rangle}{3\Mp^2H^2}\simeq\frac{\alpha A_R^2(2+s\Omega_\chi)}{12H^2},
\end{equation}
using the canonical scalaron at $F\simeq1$. For the two tested couplings, the amplitudes extrapolated with WKB give fractional mean-Hubble shifts of $6.578\times10^{-9}$ and $1.015\times10^{-13}$ on the frozen background. The corresponding scalaron energy fractions are $1.356\times10^{-8}$ and $2.091\times10^{-13}$. Numerical phase averaging at 60-digit precision, together with a solution of the frozen mean constraint, reproduces the leading shift to within $4\times10^{-9}$ in relative terms for the larger coupling. This agreement is internal to the leading phase ansatz. Testing nonlinear averaging would require a separate evolution. We hold the scalar data fixed here, so the shift does not predict $H_0$ after endpoint refitting. A validated calculation would also need continuous mean evolution, the induced changes in the other sources, and bounds on the associated likelihood errors.

\subsection{Evolution of the averaged scalar feedback}\label{sec:mean}
We extend the frozen calculation by evolving the leading scalar response to the mode from a matching redshift to today. The construction uses the GR reference and defines an averaged family; it does not continue the early nonlinear scalar trajectory exactly. We take the transported curvature amplitude from the preceding fit. Writing $h=A_R/(6H\omega)$, $B=B_{\mathrm{osc}}/H_{\mathrm{ref}}^2$, and $m=\delta\bar E/E$, we write the averaged constraint and scalar response as
\begin{align}
m&=\frac{v\delta v/3+\delta U+B}{2E^2(1+s\Omega_\chi)},\\
(\delta v)_N&=-3\delta v+\frac{3\lambda}{E}
\left(\delta U-Um+\frac{Uh^2}{2}\right),\\
(\delta U)_N&=-\frac{\lambda}{E}
\left(v\delta U+U\delta v-Uvm+\frac{Uvh^2}{2}\right).
\end{align}
The phase average of $1/H$ generates the $h^2/2$ terms at leading order. Correlations with rapid scalar responses are suppressed by higher orders in $H/\omega$ and are omitted. We evolve a forced solution and a homogeneous scalar-amplitude solution, choosing the initial velocity variation from the same approximate regular radiation-era relation as before. We combine the responses so that $m(0)=0$. The endpoint normalization is thus a boundary choice.

Before refitting, the integrated mean shift at today is $6.6600\times10^{-9}$ for $\ar=10^{-15}$; the frozen calculation gives $6.5783\times10^{-9}$. To impose the normalization, we shift the initial potential density by $-3.9967\times10^{-8}$ at $z=1000$. The normalized mean correction reaches a maximum of $1.7877\times10^{-8}$ on $0\le z\le3$. For $\ar=10^{-16}$, the corresponding values are $1.0274\times10^{-13}$, $-6.1657\times10^{-13}$, and $2.7579\times10^{-13}$. Applying the same averaged prescription from $z=3000$ gives practically identical normalized profiles at low redshift. This agreement supports the internal consistency of the matching procedure, but does not show that the result is independent of the exact scalaron initial conditions.

We check the linearization with an independently implemented nonlinear averaged closure. Its baseline-subtracted constraint is solved in stable difference variables, using the exact sinusoidal inverse-Hubble average $(1-h^2)^{-1/2}$. We adjust the potential amplitude by shooting to impose the same mean endpoint normalization. For the larger coupling, the maximum difference between the normalized low-redshift mean Hubble corrections is $1.45\times10^{-16}$, or $8.1\times10^{-9}$ relative to the correction itself. This agreement checks the two averaged descriptions, but does not validate them against the exact metric quadratic dynamics.
\subsubsection{Nonlinear checks in separate time windows}
We perform separate nonlinear runs near $z=1000,100,3,0$ for each coupling, selecting intervals of nominally twelve cycles. Those close to the present terminate at $z=0$. The changing frequency brings the recorded phase counts to approximately 12.00--13.21. To initialize each run, we add the transported curvature oscillation to the prescribed mean scalar data and solve the exact Friedmann constraint for the starting Hubble rate. Across the eight runs, the largest difference in $H$ between Radau and BDF with tighter tolerances is $9.91\times10^{-12}$. We find constraint residuals below $1.67\times10^{-14}$ and $F$ positive throughout.

The independently fitted curvature and Hubble amplitudes satisfy the leading relation to within $3.45\times10^{-6}$. The largest relative waveform residual is $2.41\times10^{-5}$ in curvature and $8.98\times10^{-5}$ in the Hubble rate. For these fits, the phase is calculated from the integrated $H$ of the individual nonlinear window. They check local amplitude transport and the curvature-to-Hubble relation, but provide no independent test of the phase accumulated across the full history. Separate initializations also mean that the windows do not form a continuous exact solution or constrain the accumulated error. Full-interval nonlinear accuracy has yet to be established.

\section{Matching prescriptions and numerical settings}\label{app:initialdata}
This appendix gives the reference normalization, curvature-matching prescriptions, and solver tolerances. They specify the initial-value problems behind the local integrations and show how those calculations differ from the reduced solutions and mode transport.
\subsection{GR reference and scalar normalization}
We describe the GR reference using $x=\dot\phi/(\sqrt6\Mp H)$ and $y=\sqrt V/(\sqrt3\Mp H)$, with $E=e^\ell$. Their evolution equations are
\begin{align}
x_N&=-3x+\sqrt{3/2}\lambda y^2-\epsilon_g x,\\
y_N&=-\sqrt{3/2}\lambda xy-\epsilon_g y,\qquad \ell_N=\epsilon_g,\\
\epsilon_g&=-\frac{3+3x^2-3y^2-3\Omega_\chi+\Omega_r}{2(1+s\Omega_\chi)}.
\end{align}
Starting at $z=2\times10^5$, we set $x_i=\sqrt6\lambda y_i^2/10$ and calculate $E_i$ from the constraint. Shooting in $y_i$ imposes $x_0^2+y_0^2=(1-f_\chi)(1-\Omega_{m,\mathrm{ref}}-\Omega_{r,\mathrm{ref}})$, which fixes the reference normalization. We use DOP853 with a relative tolerance of $2\times10^{-10}$, componentwise absolute tolerances of $(10^{-35},10^{-28},10^{-12})$ for $(x,y,\ell)$, and a maximum step of $0.025$ in $N$. The formal responses initialized at $z=10^4$ and $z=2\times10^5$ both use this reference.

We construct the response from a forced solution and a homogeneous scalar-amplitude solution, with initial derivatives $(v_{1f},U_{1f})$ and $(v_{1h},U_{1h})$, respectively. The code sets $U_{1f}=0$ and $U_{1h}=1$, then obtains the corresponding velocity derivatives by differentiating $v_i\simeq3\lambda U_i/(5E_i)$ together with the linearized constraint. We fix the coefficient combining the two solutions by imposing $e_1(0)=0$. This procedure defines the boundary family for the response, but does not establish an all-order scalaron attractor.
\subsection{Finite-coupling windows and reduced evolution}
For the local response tests using the original equations, let $e_1(N)$ denote the dimensional Hubble-response coefficient and define $\epsilon_g=(E_g)_N/E_g$. The corresponding geometric curvature variations are
\begin{align}
r_1&=6E_g[(4+\epsilon_g)e_1+(e_1)_N],\\
q_1&=e_1q_g/E_g+E_g(r_1)_N.
\end{align}
We take the initial values to be $(E_g+\ar e_1,r_g+\ar r_1,q_g+\ar q_1,0,v_g+\ar v_1,U_g+\ar U_1)$. Derivatives are evaluated with four-step, one-sided, fourth-order stencils of step $10^{-3}$, using only points within the computed interval. We repeat the calculation with a step of $5\times10^{-4}$ as a check. For the local family starting at $z_i=3$, we adjust $U_i$ to satisfy the constraint and retune the response amplitude by shooting until $E(0)=1$. In the radiation-window run, we start at $z_i=10^4$ with the scalar data held fixed. The constraint determines $E_i$, and we integrate down to $z=1000$. The initial-data matching prescription is therefore different in these two integrations.

For the unscaled late-time variables $(E,r,q,\varphi,v,U)$, we use Radau with a relative tolerance of $2\times10^{-10}$, absolute tolerances of $(10^{-10},10^{-8},10^{-7},10^{-12},10^{-11},10^{-11})$, and a maximum step of $0.008$ in $N$. The benchmark comparison uses BDF with a relative tolerance of $2\times10^{-11}$, absolute tolerances of $(10^{-11},10^{-9},10^{-8},10^{-13},10^{-12},10^{-12})$, and a maximum step of $0.002$. In the fixed-scalar test, the Radau maximum step is $0.004$; for both algorithms, the componentwise absolute tolerances are the relative tolerance multiplied by $(0.1,10,100,0.01,0.1,0.1)$. The normalized local-response tests use a maximum step of $0.003$.

For the radiation-window integrations, each state component is first divided by $\max(|Y_i|,1)$. We set the relative tolerance to $2\times10^{-10}$ for Radau and $2\times10^{-11}$ for BDF. In the scaled variables, both use an absolute tolerance equal to $0.01$ times their relative tolerance, a maximum $N$ step of $0.003$, and a complex-step Jacobian.

The reduced closure, mode fits, averaged feedback, and distributed windows can each be run separately from the release, which records their starting states. In every distributed window, we prescribe the mean scalar data and transported curvature amplitude before solving the constraint for $E_i$. The phase is then integrated using the Hubble rate from the original-equation solution in the same window. An independent calculation of the phase over the entire history is still needed.
\subsection{Reproduction and accuracy interpretation}
The command \texttt{python reproduce.py --list} lists the tasks; \texttt{python reproduce.py --all} runs them in sequence. Package versions, initial data, data hashes, and numerical results are recorded in the release. The two figures are reproduced by the designated plotting tasks. Software versions and hardware can affect solver differences at machine precision; the relevant check is agreement within the quoted tolerances, rather than bitwise equality.

Table~\ref{tab:slow-defects} reports residuals of the original equations at sampled points, rather than estimates of the solution error. In particular, the trace defect of about $1.17\%$ at $\ar=10^{-15}$ cannot establish the accuracy of distance corrections that are many orders of magnitude smaller. The covariance envelopes in Sec.~\ref{sec:distance} bound only the specified oscillatory ansatz. Neither the local checks against the original equations nor the agreement between averaged closures provides an error bound over the full evolution.

\end{document}